# μSR-detection of the soft-mode toward the exotic magnetic ground state in bond-disordered quantum spin system IPA-Cu(Cl$_{0.35}$Br$_{0.65}$)$_3$

Takayuki Goto[1], Takao Suzuki[2], Keishi Kanada[1], Takehiro Saito[1], Akira Oosawa[1]
Isao Watanabe[2] and Hirotaka Manaka[3]

[1]*Department of Physics, Faculty of Science and Technology, Sophia University 7-1 Kioicho, Chiyodaku Tokyo 102-8554, Japan*

[2]*Advanced Meson Science Laboratory, Nishina Center, RIKEN, 2-1 Hirosawa, Wako, Saitama 351-0198, Japan*

[3]*Graduate School of Science and Engineering, Kagoshima University Korimoto, Kagoshima 890-0065, Japan*

The spin-gap system with a bond-disorder (CH$_3$)$_2$CHNH$_3$-Cu(Cl$_x$Br$_{1-x}$)$_3$ is reported by experiments on macroscopic quantities of the uniform susceptibility and the specific heat to show a magnetic order at low temperatures when the value of *x* is within the limited region between 0.44 and 0.87, and otherwise a gapped behavior. We have investigated microscopically the ground state of the sample with *x*=0.35 by muon spin relaxation (μSR) to observe that the frequency spectrum of the spin fluctuation tends to concentrate around zero energy as lowering temperatures. This behavior was interpreted as the soft-mode toward the possible magnetic phase transition at absolute zero.

**1. Introduction**

Disorder often introduces a non-trivial effect into the quantum spin systems. In high-$T_C$ cuprates, the uniform Néel order in the parent compound is rapidly destroyed by doping a small fraction of holes; this is known as a frustration effect.[1,2,3] Conversely in spin Peierls systems, the Néel state is stabilized by substituting a small amount of non-magnetic impurities, which destroy singlet dimers and create unpaired spins.[4,5] Furthermore, there is a theoretical prediction of the appearance of an exotic ground state Bose-glass in disordered Boson systems.[6,7] A disordered spin system is one of the candidate that bears the ground state of the Bose-glass. Though possibility of the experimental detection of the Bose-glass phase by either macroscopic[8,9,10] or microscopic[11,12] probes was repeatedly suggested in the solid-solution of different two spin-gap systems, details still remain obscure.

The compounds IPA-CuCl$_3$ and IPA-CuBr$_3$, where IPA stands for isopropylammonium (CH$_3$)$_2$CHNH$_3$, are spin gap systems with an excitation gap $\Delta$=14 and 98 K, respectively.[14,15] IPA-CuCl$_3$ was formerly reported to be $S$=1/2 ferromagnetic-antiferromagnetic bond-alternating chain along $c$-axis[16], and recently, one of authors (H. M) and his group have revealed by inelastic neutron scattering experiments that it is a two-legged spin ladder system with the strongly-coupled ferromagnetic rungs, which are aligned on the leg along $a$-axis.[14] Therefore, the system can be considered to be an effective $S$=1 antiferromagnetic chain, and hence the composite Haldane system, which is expected to have a gapped singlet ground state.[17,18-20] Br-system has so far been reported to be $S$=1/2 antiferromagnetic-antiferromagnetic bond-alternating chain[15].

Intensive investigation on specific heat and uniform magnetization of the solid solution of the two systems with various values of Cl concentration ratio $x$ by one of the authors (H. M.) have shown that there are two quantum critical points (QCP) at $x$=0.44 and 0.87 and that a magnetic order occurs only when $x$ is in between these two values.[21,22] Nakamura theoretically treated this system by the quantum Monte-Carlo method, and explained that the appearance of the uniform Néel state in this bond-disordered system, which, from the classical view-point, favors a glass-like state, is entirely due to quantum effects.[23] He also claims that a QCP exists only in Cl-rich phase, but not in Br-rich phase. Fisher has theoretically studied the disordered lattice boson system with random on-site chemical potential to predict the existence of an exotic ground state Bose-glass.[6] The phase exists only at absolute zero, and is characterized most by the property that bosons are localized though they are massless. If we map the boson particle onto the triplet, and the vacuum onto the singlet, the present spin system is well-described as the disordered boson system, which bears possibility to be in the Bose-glass phase at absolute zero. Note that if one changes the boson number density by tuning the chemical potential, which in spin language, corresponds to the applied field, the Bose-Einstein condensation of boson takes place at low temperatures, as is well known as magnon-BEC.[25,26]

The motivation of our study is to investigate microscopically the ground state of the mixture of these two non magnetic compounds. The mixture can be considered as the quantum spin system with disordered bonds. The disorder in this system differs qualitatively from the Mg-doped spin Peierls systems such as $CuGeO_3$ [4,5] in that there are no unpaired spins in the present system, but only a random modulation of the bonds.

In our previous papers, we reported μSR and NMR results on Cl-rich samples with $x$=0.85, 0.88 and 0.95. The first one with $x$=0.85, the value of which resides within the two QCPs, undergoes static and uniform antiferromagnetic order at $T_N$=11.65K, which was demonstrated by clear muon spin rotation, NMR-peak splitting, and critical divergence of the nuclear spin-lattice relaxation rate $T_1^{-1}$.[27-29] As for the other two with $x$=0.95 and 0.88, which are outside of the two QCPs, and hence were believed to be gapped, we confirmed that though they do not show a magnetic order down to the dilution temperature region[31], their spin relaxation rate stay at anomalously high value[27,30]. Though this fact at first glance suggests that their ground state is magnetic, many μSR reports performed so far on other quantum spin systems warn that it is dangerous to judge the ground state to be magnetic or non magnetic only from the magnitude of the muon spin relaxation rate [32,33,34,35].

In order to determine the ground state of quantum spin systems, we have developed a simple way to obtain a spin fluctuation spectrum from μSR experiments under longitudinal field (LF). In this paper, we present an application of this method to a Br-rich sample with $x$=0.35, which was reported by macroscopic experiments to be the singlet dimer phase, the same as $x$=0, and have a finite spin excitation energy[21,22]

## 2. Experimental

Single crystals of $(CH_3)_2CHNH_3Cu(Cl_xBr_{1-x})_3$ with $x$=0.35 have been grown by an evaporation method from isopropylalcohol and methanol solution of $(CH_3)_2CHNH_2 \cdot HX$ and $CuX_2$ with X=Cl, Br.[15,16,21,22] The Cl concentration ratio $x$ in the crystals was determined by the inductively coupled plasma atomic emission spectrometry (ICP-AES) method. The external shape of the crystals was a rectangular-solid. Three single crystals with total volume around 400 mm$^3$ were used for μSR experiments. They were attached by an Apiezon N grease to a silver plate with a purity of five nines,

ver. 080706

connected at the bottom of the cryostat.

Measurements of μSR were carried out at the Riken-RAL Muon Facility in the U. K. using a spin-polarized pulsed surface-muon ($\mu^+$) beam with a momentum of 27 MeV/c. The decay of the muon spin polarization is described by the asymmetry parameter $A(t)$, which is proportional to the spin polarization of the muon ensemble and is obtained from the ratio of numbers of muon-events counted by the forward and backward counters. The incident muon beam was directed parallel with $b^*$-axis, which was perpendicular to one of the three outer planes with the largest area, so-called C-plane.[16,31] The sample temperature was controlled with an Oxford $^3$He cryostat in the range of 0.3-14 K. The typical temperature stability during measurements was ±10 mK above 1.5 K, and ±2 mK below 1.5 K.

3. Results

Typical relaxation curves of muon spin polarization under zero and finite longitudinal fields (LFs) are shown in Figs. 1 and 2. As the temperature was lowered, one can see that the relaxation becomes faster, and that the shape of the short time region of the curves changed from Gaussian-type to Lorentzian-type. The relaxation over the entire temperature and field region was described by a function containing two components expressed as

$$A(t) = A_1 G_{KT}(t,\Delta)\exp(-\lambda_1 t) + A_2\exp(-\lambda_2 t) \qquad (1)$$

where $G_{KT}$ is the Kubo-Toyabe function with parameters of a static field distribution width at the muon site $\Delta$, relaxation rates $\lambda_1$ and $\lambda_2$, and component amplitudes $A_1$ and $A_2$. No muon spin rotation due to a static field by a magnetically ordered phase was observed in any temperature or field region. A slight wiggling in relaxation curves of 5 K at around 5 µs resulted from the nuclear contribution[36], and was clearly reproduced by the adapted fitting function, indicating the validity of the fit. This wiggling was obscured by the fast relaxation under the zero field (ZF) condition at 0.3 K, where the two components of the relaxation $\lambda_1$ and $\lambda_2$ became very large. The magnitude of $\lambda_2$ always exceeded $\lambda_1$ by more than one order of magnitude. The amplitude fraction of $\lambda_2$, the faster relaxation, was around $A_2/(A_1+A_2) \approx 0.1$ at 8K, increased with decreasing temperature, and reached 0.4 at 0.3 K.

In determining these parameters at zero field, we first set all of them free and performed fitting to find that $\Delta$ distributes randomly within a region 0.2-0.27, and was considered to be the nuclear contribution. Next, we fixed $\Delta$ to be the averaged value 0.23 ($\mu s^{-1}$) and performed all the fitting again to obtain the other parameters.

The longitudinal-field (LF) decoupling at the lowest temperature 0.3 K is shown in Fig. 3. In determining the parameters at finite fields, we first set the fraction of each two components $A_1$ and $A_2$ as free and confirmed that the fraction does not vary against LF below 50 Oe. Above 50 Oe, where the slower relaxation rate $\lambda_1$ becomes negligibly small and hence contribute little to the relaxation curves of measured time range, we fixed the fraction to the low field values and $\lambda_1$ to zero, and assumed that the fractions does not change from the value taken at low fields. Observed relaxation curves above 50 Oe showed a good fit with the function based on this assumption.

The LF-dependence of the two relaxation rates, which maps the Fourier spectrum of spin fluctuation was analyzed by the Lorentz function, commonly referred as the Redfield equation[13]. This function represents the behavior of classical localized spins, and is expressed as

$$\lambda(H_{LF}) = \frac{2(\gamma_\mu \delta H_{loc})^2 \tau_C}{1 + (\gamma_\mu H_{LF} \tau_C)^2} \qquad (2)$$

where $\delta H_{loc}$, $\tau_c$ and $\gamma_\mu$ are the local-field fluctuation amplitude, the characteristic fluctuation time constant, and the muon spin gyro-magnetic ratio (13.5534 kHz/Oe). By fitting the observed $\lambda_1(H)$ and $\lambda_2(H)$ to the above function, we obtained parameters to be $\delta H_{loc}$=7.0(6) Oe and $\tau_c$=7.6(5) μs for the $A_1$ component, and 39(4) Oe and 1.8(2) μs for the $A_2$ component, respectively.

As shown in Fig. 3, a good fit was achieved for $\lambda_1$ below 50 Oe, above which $\lambda_1$ becomes negligibly small. However, as for $\lambda_2$, a satisfying fit was not achieved when the fitting function is applied to the entire experimental field range up to 3900 Oe. Therefore, we limited a fitting range below 50 Oe to obtain parameters shown above. The slight deviation from the Lorentz curve above 100 Oe indicates that the spin fluctuation spectrum is not expressed by a unique characteristic time and hence that the $\lambda_2$ component include an additional fluctuation with characteristic time still shorter than 1.8 μs. This may be brought by the finite temperature effect. In higher temperature region, observed data cannot be expressed by Eq. (2), supporting our speculation.

The fluctuation frequencies of the two components were different beyond fitting errors, indicating that the origin of the driving force in the two components are different. This fact denies the possibility of the two muon sites model for the appearance of the two components as will be discussed below. Note that the curvature of the two decoupling curves is significantly different from each other even below 30 Oe, which is expected to be too lower to affect the spin state. This assures that the difference in the frequencies for the two components is not an artifact brought by the applied LF.

ver. 080706

The temperature dependence of the two relaxation rates under various LFs is shown in Fig. 4. Under a low LF of 20 Oe, both $\lambda_1$ and $\lambda_2$ increased with decreasing temperature. The larger component $\lambda_2$ reached 0.8 $\mu s^{-1}$ at 0.3K. When the LF was raised to 100 Oe, the increase in the relaxation rates quenched at 3K, below which both $\lambda_1$ and $\lambda_2$ decreased with decreasing temperature. The slower component $\lambda_1$ was readily suppressed completely by an LF of 300 Oe, above which only the $\lambda_2$ component survived. In still higher LFs, $\lambda_2$ decreased with decreasing temperature from the highest measured temperature 8 K. The temperature at which $\lambda_2$ shows maximum increases monotonically with increasing LF.

3. Discussion

The absence of the muon spin rotation and the fact that the decoupling data are well described by a classical Lorentzian spectrum indicate that there is no long range magnetic order down to 0.3 K, where the electron spins fluctuate paramagnetically. This observation is consistent with the macroscopic experiments that have reported an absence of magnetic order at $x$ =0.35.[21] However, the large values of the two local fields $\delta H_{loc}$=38.5 and 7.7 Oe, which were determined by the LF-detailed decoupling experiments indicate that the system is not uniform and that the ground state is magnetic. This is in a clear contrast to the report of macroscopic experiments [21,22], which claims a uniform and non-magnetic gapped ground state.

The appearance of the magnetic ground state detected only by a microscopic probe of µSR experiments indicates two possibilities. One is a critical enhancement of the generalized susceptibility $\chi(q)$ with a non-zero $q$-vector, which is characteristic to the

-8-

order phase between $x$=0.44-0.87, though its value, together with its concentration dependence are not determined yet in the present system.  This kind of phenomenon was reported in high-$T_C$ cuprates, where the staggered susceptibility is critically enhanced with decreasing the hole concentration toward the QCP which divides the superconducting and antiferromagnetic regions.[38]  The other possibility is emergence of an exotic magnetic ground state such as the Bose-glass phase, which shows no magnetic order at finite temperature[6].  However, we do not deny the still other possibility of the magnetic ground state other than the Bose-glass, and with the finite transition temperature much lower than $^3$He temperature region.  We will discuss this point again later with the results of the decoupling at the various temperatures.

Next, we discuss the origin of the two components with different fluctuation time constants.  First, the region with larger relaxation rate $\lambda_2$ is considered to be finite-sized islands, to which active spins are restricted.  They should be antiferromagnetically-correlated rather than be isolated from each other.  This is because the system does not show a large Curie-term in dc magnetization measurements at low temperatures, though the volume fraction of the magnetic component reached 0.4. Furthermore, if isolated spins were present and acted as relaxation centers, the relaxation curve would have to be a stretched-exponential type function, as is often reported in dilute magnetic alloys[39,40] or doped spin-Peierls systems[41,42].  In fact, the observed relaxation always obeyed a simple-exponential-type function over the entire range of experimental temperatures and fields, as stated above, assuring that the island picture is plausible.

The region with the small relaxation rate $\lambda_1$ is assigned to the singlet sea, which is inherent to the gapped phase in the end member compounds IPA-CuBr$_3$ or IPA-CuCl$_3$, because the magnitude of the relaxation rate $\lambda_1$ at low temperature is as small as that in IPA-CuCl$_3$[43]. However, the characteristic fluctuation frequency of the spins at the lowest temperature is as slow as 0.14 MHz, much smaller than 99 MHz in IPA-CuCl$_3$ at 2 K [43]. This means that the spin fluctuation in the singlet sea is significantly slowed down by the antiferromagnetic spin correlation in the islands.

This phase separation is in microscopic length scale, and considered to take place due to the bond-randomness effect. The upper and lower bounds of the island size is determined respectively by the two conditions that the number of spins in islands is not large enough to show thermodynamic phase transition, and not too small to allow an emergence of Curie term in the uniform susceptibility.

Our recent NMR study on the Cl-rich sample $x=0.88$, that was reported by macroscopic experiments to have a gapped ground state, has also shown that the system shows a microscopic phase separation of this island picture.[30] In our previous µSR paper on samples $x=0.95$ and 0.85, which were reported by macroscopic experiments to be gapped and gapless, respectively, we observed the two relaxation components in muon time spectra, and interpreted them as either from two muon stopping sites or the microscopic phase separation[27]. The present µSR results support the latter possibility in ref 27.

Note that a possibility of macroscopic phase separation in this system is readily ruled out for two reasons. First, all the sample crystals were free from spurious Curie terms in a uniform susceptibility [21,22]. The second reason is the absence of long-range order. This fact assures that the sample was not contaminated by the magnetically ordered phase between $x=0.44–0.87$, which, if exists, must show a magnetic phase transition between 13 and 17 K [21]. From these observations, we can conclude that the system intrinsically consists of two parts, one with strong magnetic fluctuation corresponding to $\lambda_2$ and the other with extremely weak magnetic fluctuation corresponding to $\lambda_1$, and both of which stay paramagnetic down to 0.3 K.

Next, we demonstrate the development of the soft mode toward the possible magnetic phase locating at absolute zero or finite temperature much lower than $^3$He temperature region, from the temperature dependence of the relaxation rates under the various LFs shown in Fig. 4. In the configuration with an LF high enough to collapse the quasi-static nuclear contribution, that is, higher than a few tens Oe, the muon relaxation is driven only by the fluctuating local field which is directed perpendicular to the muon spin polarization and has the characteristic frequency which is the same with the Larmor frequency of muon spins $\omega_{LF} = \gamma_\mu H_{LF}$. Therefore, we can see that Fig. 4 shows the temperature dependence of the Fourier component with frequency $\omega_{LF}$ in the fluctuating local field. The observed temperature dependence indicates that the amplitude with lower frequency increased as temperature was lowered, while that for higher frequencies decreased. This vividly maps the change in the frequency spectra of the magnetic fluctuation with the temperature. At 8K, the spectrum was nearly white, which can easily be seen in the cross sectional view of Fig. 4, and as temperature was lowered, the low frequency part was gradually weighed, indicating an emergence of the critical slowing down of the characteristic frequency of the magnetic fluctuation. This behavior is a typical soft mode, which is generally seen around the second-order phase transition. The appearance of the soft mode indicates the existence of the phase transition at either the one located at absolute zero such as Bose-glass, or a magnetic order with a transition temperature much lower than 0.3 K.

The existence of the soft mode is direct evidence for the magnetic ground state in the present phase.  In above, we have pointed out the two possibilities for the ground state of this system; one is the critical enhancement around QCP and the other is the exotic ground state such as Bose-glass.  We can conclude that the latter is plausible in the present system.  This conclusion is consistent with the theoretical investigations. Nakamura showed by QMC calculation that the antiferromagnetic dimer state, which corresponds to IPA-CuBr$_3$, is remarkably instable against the bond disorder and easily brought to a critical state between gapped and gapless states[23].  This makes a clear contrast to the Cl-rich region, where the Haldane state is robust against bond disorder as shown by Hida with the density matrix renormalization group calculation[24].

Next, we test the possibility of the Bose-glass for the candidate for the ground state of the present case.  According the Fisher's theoretical prediction on the Bose-glass, the phase is located at absolute zero, and has finite compressibility, which in the spin language corresponds to the finite uniform susceptibility.  In fact, a small but finite uniform susceptibility at low fields was reported by Manaka *et al.*[21,22] in the present concentration $x$=0.35.  This will strongly support our speculation.

Fisher predicted still other two properties of the Bose-glass.  One is the localization of the triplet and the other the appearance of the characteristic curvature in the phase boundary between Bose-glass and BEC phase.  The evidence of the localization has actually been reported in the other disordered quantum spin system (Tl,K)CuCl$_3$, which is believed to have the ground state of the Bose-glass[8,9,10], as a discrepancy between the temperature dependence of the uniform and the local magnetization, the latter of which is probed by NMR shift[11].  In the present case, our tentative NMR study shows different temperature dependence between NMR shift and the uniform susceptibility.  This also supports the Bose-glass picture.  In order to test the Fisher's latter prediction on the phase boundary between Bose-glass and BEC, an investigation on the field-induced magnetic order is necessary, and is now in progress. In order to confirm that this soft mode continues to slow down toward absolute zero or the magnetic order exists at the finite temperature much lower than $^3$He region, an experiment in the dilution temperature range is also indispensable.  This measurement is also in progress.

Finally, we note the merit of studying the soft mode by muon techniques.  First, this method allows one to capture the soft mode if one is ignorant of the magnetic $q$-vector for the possible ordered phase, or even if the relaxation does not corresponds to a well-defined collective mode in $q$-space.  This is simply because the muon spin is a local or point-like probe, which detects magnetic instability with any spatial periodicity or even without it.  Next, μSR allows one to trace the energy spectra with the resolution less than 1 K toward the critical point, while the neutron techniques, by which the soft mode is conventionally studied, is often interrupted halfway toward the critical point because of the instrumental energy resolution limit or of the disturbance from the incoherent scattering.

**Summary**

Muon spin relaxation in the bond-disordered quantum spin system IPA-Cu(Cl$_{0.35}$Br$_{0.65}$)$_3$, which was reported by experiments on macroscopic quantities to have a gapped ground state, was investigated.  The system exhibited two relaxation components with different amplitudes and frequencies, indicating the microscopic inhomogeneity arising from the bond randomness effect.  Both the two relaxation rates showed steep increases with lowering temperature down to 0.3 K, indicating that the ground state is gapless. The existence of the soft mode toward the magnetic ground state is successfully probed by the temperature dependence of the characteristic frequency of the magnetic fluctuation.

**Acknowledgements**

The authors are grateful for kind and valuable discussion with Prof. Tota Nakamura. This work was supported by KEK-MSL Inter-University Program for Overseas Muon Facilities and by a Grant-in-Aid for Scientific Research on the Priority Area "High Magnetic Field Spin Science in 100T" from MEXT.

ver. 080706

ver. 080706

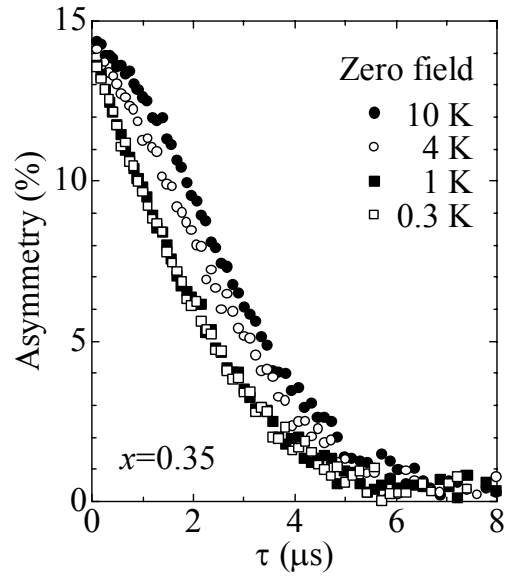

Fig. 1 Typical relaxation curves at various temperatures under zero field.

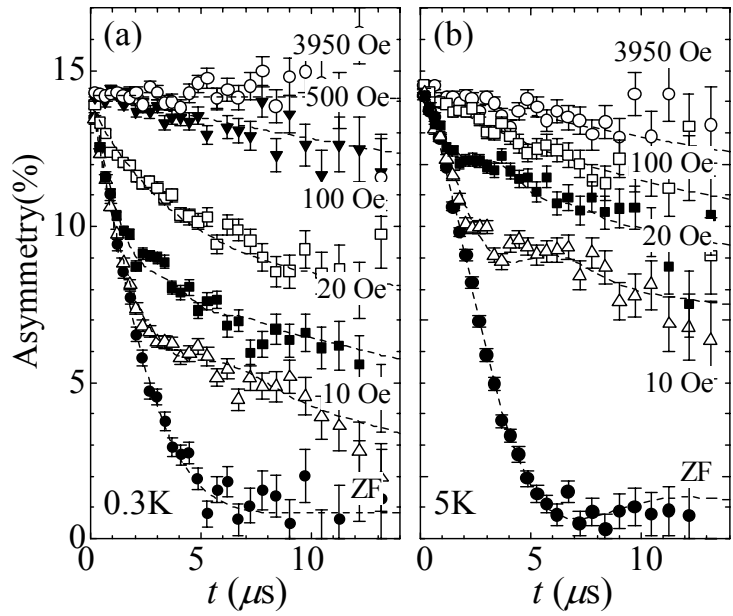

Fig. 2 Typical relaxation curves under various magnetic fields at (a) 0.3 K and (b) 5 K. Dashed curves are the fitting function Eq. (1).

-17-

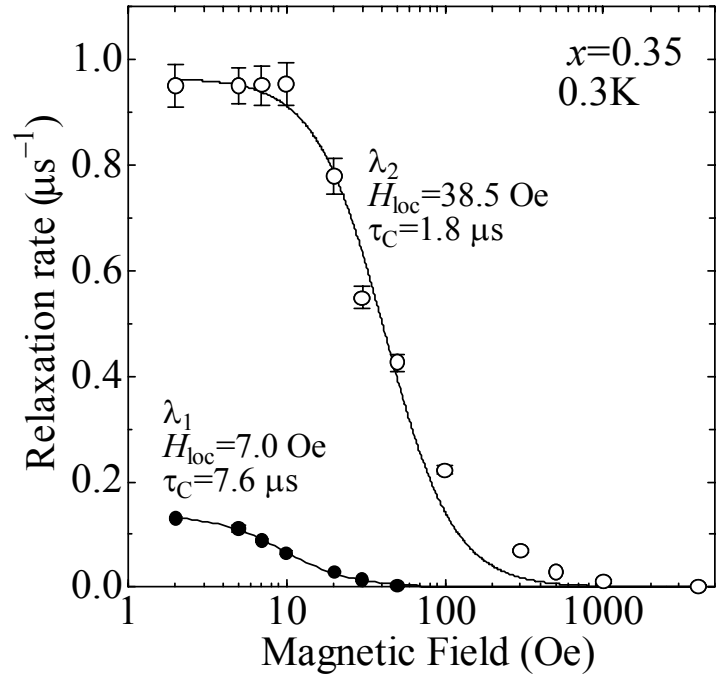

Fig. 3 Decoupling with longitudinal fields up to 3900 Oe at 0.3K.   Solid curves are the Redfield function of Eq. (2).   Fitted error bars for $\lambda_1$ are around 0.001-0.005 and hence are hidden behind symbols.

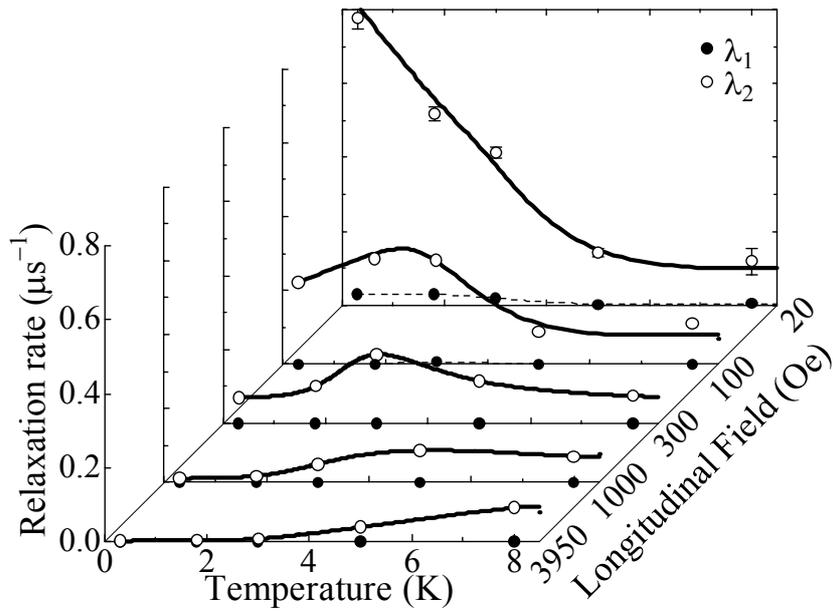

Fig. 4 Temperature dependence of relaxation rates $\lambda_1, \lambda_2$ corresponding to the two

components under various longitudinal fields.  Solid and dashed curves are guides for the eye.